\documentclass[12pt]{article}
\usepackage{amssymb,latexsym}
\usepackage{amsmath}

\columnwidth\textwidth

\begin{document}

\newcommand{\be}{\begin{equation}}
\newcommand{\ee}{\end{equation}}
\newcommand{\beann}{\begin{eqnarray*}}
\newcommand{\eeann}{\end{eqnarray*}}
\newcommand{\bea}{\begin{eqnarray}}
\newcommand{\eea}{\end{eqnarray}}
\newcommand{\nn}{\nonumber}
\newtheorem{df}{Definition}
\newtheorem{thm}{Theorem}
\newtheorem{lem}{Lemma}

\begin{titlepage}

\noindent

\vspace*{1cm}
\begin{center}
{\LARGE Can differently prepared mixed states be
distinguished?}

\vspace{2cm}

J. Tolar \\
Department of Physics \\ Faculty of Nuclear Sciences and
Physical Engineering\\ Czech Technical University \\
B\v{r}ehov\'{a} 7, CZ-115 19 Prague 1, Czech Republic \\
jiri.tolar@fjfi.cvut.cz \\
\vspace{0.5cm}

P. H\'{a}j\'{\i}\v{c}ek \\
Institute for Theoretical Physics \\
University of Bern \\
Sidlerstrasse 5, CH-3012 Bern, Switzerland
\\ hajicek@itp.unibe.ch \\
\vspace{1.5cm}

December 2005 \\ \vspace*{1cm}

\nopagebreak[4]

\begin{abstract}
A measurement has been proposed by B. d'Espagnat that would
distinguish from one another some ensembles that are differently
prepared but correspond to the same density matrix. Here, the idea
is modified so that it becomes applicable to much more general
situations. The method is illustrated in simple examples. Some
matter of concern might then be that information could be
transmitted by methods based on our idea in the EPR kind of
experiments. A simple proof is given that this is impossible.



\end{abstract}

PACS number(s): 03.65.Ta; 03.67.-a


\end{center}

\end{titlepage}

\section{Introduction}
In chapters 7 and 8 of his book \cite{Esp}, B. d'Espagnat discusses
the quantum mechanical notion of mixed states in terms of ensembles
related to preparation of such states. He presents examples of mixed
ensembles prepared by mixing a great number $N$ of polarized spin
1/2 particles in several different ways, all of them leading,
however, to the unpolarized beam.

It is a well known fact that the values of all measured quantities
are uniquely determined if the density operator of the system is
given. Hence there is no operational method to distinguish between
physically different realizations of the ensembles, provided they
lead to ``states'' with the same density operator. In his examples
$\rho= \frac{1}{2}\text{Id}(2)$, where Id$(2)$ is the $2\times 2$
unit matrix.

In order to test the difference between the ensembles
 experimentally, B. d'Espagnat presents on p.\ 122 of
 \cite{Esp} a proposal:
`Can we produce such a difference?  Yes, we can, provided we treat
these ``ensembles'' as what, after all, they physically are, that
is, as systems of (noninteracting) particles.' The aim of this paper
is to generalize and expand this proposal. Our motivation and
concern is the well-known fact (see, e.g., \cite{peres}, p.\ 170)
that, in the EPR experiment, superluminal signals could be sent if
such differences were measurable in general. However, we show in
Sect. 3 that applying to the context of the EPR experiment the here
considered generalization of d'Espagnat's method does not lead to
the possibility of transmitting information.

\section{Ensembles and mixed states}
In quantum mechanics a separable Hilbert space $\cal H$ is associated to each
physical system. Its states are represented by the density operators. We
assume that no superselection rules are operating, hence the correspondence
between states and density operators is one--to--one.  The family $\cal S$ of
density operators is naturally closed under convex combination. The projectors
onto one--dimensional subspaces belong to $\cal S$ and represent the pure
states of the system. They are the extremal elements of $\cal S$. For more
details, see \cite{JvN}.

In the notation of \cite{Esp} a mixed ensemble $\hat{E}$ consisting of a large
number $N=\sum_{\alpha=1}^{r}N_{\alpha}$ particles of which $N_{\alpha}$,
$\alpha = 1,\dots,r$, belong to pure states $|\phi_{\alpha}>$, is described by
the density operator
$$
\rho=\sum_{\alpha=1}^{r}\frac{N_{\alpha}}{N}
          |\phi_{\alpha}\rangle \langle\phi_{\alpha}|\ .
$$
If $|\phi_\alpha\rangle$ are pairwise orthogonal, then the decomposition of
$\rho$ is unique if and only if there is no degeneracy, i.e. $N_{\alpha} \neq
N_{\beta}$ for $\alpha \neq \beta$. But if there is degeneracy, or if the pure
states $|\phi_\alpha\rangle$ are not pairwise orthogonal, then there are
infinitely many convex decompositions of $\rho$.

The ensembles I and II are described on p.\ 121 of \cite{Esp}: `Let us
consider two methods that can be used to prepare an unpolarized beam of spin
1/2 particles.  Method I consists in mixing, by means of suitable magnets, two
fully polarized beams of equal intensity, one polarized along the $Oz$
direction, the other along the opposite direction. Method II is identical to
method I except that the $\pm Ox$ directions replace the $\pm Oz$ directions.'

Our attitude is operational, based on the usual statistical
interpretation of quantum mechanics. Thus the non-unique
decomposability of mixtures tells us that the prepared state cannot
be physically distinguished from results of infinitely many other
preparation procedures.  For instance, a state obtained by mixing up
(without phase correlations) equal numbers of spin up and spin down
particles (along $Oz$ axis, method I) cannot be physically
distinguished from a state obtained by a similar mixture of spin
left and spin right particles (along $Ox$ axis, method II).  This
fact is embodied in the formula for the expectation value of an
observable $A$ in the state $\rho$,
$$
\langle A\rangle_{\rho}= Tr(\rho A) =
     \sum_{\alpha=1}^{r}\frac{N_{\alpha}}{N}
       \langle\phi_\alpha | A | \phi_\alpha \rangle\ .
$$
For a detailed discussion of this question see \cite{beltr}.

\section{Ensembles of beams}

Let us modify the proposal of \cite{Esp} in order to apply it to EPR
situations. One should consider an ensemble of particles as being
also an ensemble of beams, i.e.\ ensemble of systems of $N$
noninteracting particles. These ensembles can in principle be
prepared along the general lines of methods I or II (only, dropping
the condition that the two fully polarized beams mentioned in
d'Espagnat's above quoted sentence should be of same intensity since
this condition is incompatible with our requirement of complete
randomness, see below). In this way, the same preparation procedure
can, if repeated sufficient number of times, be interpreted as also
preparing an $N$-particle state for any fixed positive integer $N$.
Such beams can therefore be repeatedly prepared and then subjected
to measurements. In this way, the proposal can also be formulated
without the notion of ensemble.

Let us consider a system of $N$ spin 1/2 particles with
the Hilbert space
being the $N$-fold tensor product of two--dimensional
complex vector spaces
(equipped with the standard inner products)
$$
{\cal H} = {\mathbb C}^{2} \otimes \cdots \otimes {\mathbb C}^{2}\ .
$$

To prepare the beam according to method I, we take the set
of its $2^N$ basis vectors in the form\footnote{We disregard
 quantum statistics to keep the
 argument simple. A justification is given at the end of Sect. 4.}
$$
\{ e_{i_{1}}\otimes \cdots \otimes e_{i_{N}} \}_{i_1,\dots,i_N = \pm 1}\ ,
$$
where
$$
e_{+1} = \left(  \begin{array} {c} 1 \\ 0
\end{array} \right)\ ,
\qquad
e_{-1} = \left(  \begin{array} {c} 0 \\ 1
\end{array} \right)\ ,
$$
are normalized eigenvectors of the Pauli matrix $\sigma_z$.

For method II another basis
$$
\{ f_{i_{1}} \otimes \cdots \otimes f_{i_{N}} \}_{i_1,\dots,i_N = \pm 1}
$$
is appropriate, where
$$
f_{+1} = \frac{1}{\sqrt{2}}
 \left(  \begin{array} {c} 1 \\ 1
\end{array} \right)\ ,
\qquad
f_{-1} = \frac{1}{\sqrt{2}}
\left(  \begin{array} {c} 1 \\ -1
\end{array} \right) \ ,
$$
are normalized eigenvectors of the Pauli matrix $\sigma_x$.

Now, we specify the preparation in more detail: it should satisfy
the requirement of {\em complete randomness}. A random mixture
corresponding to ensemble I of $N$-particle beams is described by
the density operator where all $2^N$ projectors on basis states
contribute with equal weights,
$$
\rho_N^{I} = \frac{1}{2^N} \sum_{i_1,\dots,i_N = \pm 1}
E_{i_{1}}\otimes \cdots \otimes E_{i_{N}}\ ,
$$
where $E_{\pm 1}$ denote orthogonal projectors on vectors
$e_{\pm 1}$.

This can also be seen in more detail using elementary
combinatorics. Imagine a chopping of a big random ensemble
of spin 1/2 particles or, equivalently, of a very long
sequence of randomly distributed $\pm 1$'s, into a sequence
of sections of length $N$ or $N$--beams. It is clear that,
in each $N$-beam, some spins (say $m$), are up and the
remaining $N-m$ are down. The number of all possible
distinct configurations of an $N$--beam with $m$ spins up
is given by the combination number
$N_{m} =  \left(  \begin{array} {c} N \\ m
\end{array} \right) $ ,
hence the total number of all possible configurations in
$N$-beams is the sum
$$
 \sum_{m=0}^{N} N_{m} = \sum_{m=0}^{N}  \left(
 \begin{array} {c} N \\ m \end{array} \right)  = 2^{N}.
$$
Now to each configuration the representative projection
operator in ${\cal H}$ is associated, e.g.
$(E_{+1})^{\otimes m} \otimes (E_{-1})^{\otimes N-m}$
corresponds to the configuration $\{+1, \dots , +1,
-1, \dots , -1 \}$. For fixed $m$, a random mixture is
obtained by assigning equal probabilities $1/N_m$ to all
$N_m$ projectors corresponding to distinct configurations
of $\pm 1$'s:
$$
 \rho^{I}_{Nm} = \frac{1}{N_m} [
(E_{+1})^{\otimes m} \otimes (E_{-1})^{\otimes N-m} +
\cdots  ].
$$
If, however, a completely random mixture including all
$m = 0, \dots, N$ is to be formed, the density matrices
$ \rho^{I}_{Nm} $ should be mixed with relative weights
$N_{m}/  \sum_{m=0}^{N} N_{m}$ proportional to numbers of
configurations contributing to each $ \rho^{I}_{Nm}$:
$$
 \rho^{I}_{N} = \sum_{m=0}^{N} \frac{N_{m} }{  \sum_{m=0}^{N} N_{m}}\rho^{I}_{Nm} =
 \frac{1}{2^N}  \sum_{m=0}^{N} \left(
 \begin{array} {c} N \\ m \end{array} \right)\rho_{Nm}^I
= \frac{1}{2^N} \sum_{i_1,\dots,i_N = \pm 1}
E_{i_{1}}\otimes \cdots \otimes E_{i_{N}}\ .
$$
But this is just the above formula for  $\rho^{I}_{N}$.

Since the result can be written as
$$
  \rho_N^{I} = \frac{1}{2}\sum_{i_{1}= \pm 1} E_{i_{1}}
 \otimes \cdots \otimes
\frac{1}{2}\sum_{i_{N}= \pm 1} E_{i_{N}}\ ,
$$
we obtain the density matrix of a completely unpolarized beam,
$$
  \rho_N^{I} = \frac{1}{2}\left(\begin{array}{ll}1&0\\0&1\end{array}\right)
  \otimes \cdots \otimes \frac{1}{2}\left(\begin{array}{ll}1&0
  \\0&1\end{array}\right)\ ,
$$
or
$$
  \rho_N^{I} =  \frac{1}{2^N}\text{Id}(2^N)
$$
in a suitable basis, where Id$(n)$ is the $n\times n$ unit matrix.

Following the same calculation with basis vectors $e_i$ replaced by $f_i$ and
projectors $E_i$ by projectors $F_i$ on eigenvectors of $\sigma_x$, the same
density matrix is obtained for case II, $ \rho_N^{II} = \rho_N^{I}$.

The $N$-particle observable of twice the $z$-component of
total spin
$$
\Sigma_{z} = \sigma_{z}\otimes I \otimes \cdots
              \otimes I + \cdots +
              I \otimes I \otimes \cdots
              \otimes \sigma_{z}
$$
has in both cases I and II the zero expectation value
$$
\langle\Sigma_{z}\rangle_N^{I,II} = Tr (\rho_N^{I,II}\Sigma_{z})= 0
$$
because $Tr \, \sigma_{z}=0$.

The dispersion $\sigma^I_{N}$ of $\Sigma_{z}$ on basis states I is
obviously zero, since the basis I consists of eigenvectors of
$\Sigma_{z}$. Let us compute it, however, on the mixed state
  $\rho_{N}^{I}$:
$$
(\sigma^I_{N})^{2} = \langle\Sigma_{z}^{2}\rangle - \langle\Sigma_{z}\rangle^2
= \frac{1}{2^N} \sum_{i_1,\dots,i_N = \pm 1}
\langle e_{i_{1}}\otimes \cdots \otimes e_{i_{N}}|
\Sigma_{z}^{2} |
 e_{i_{1}}\otimes \cdots \otimes e_{i_{N}}\rangle\ .
$$
The result is the weighted sum of eigenvalues
of $\Sigma_{z}^{2}$,
$$
(\sigma^I_{N})^{2}=
\frac{1}{2^N}\sum_{i_1,\dots,i_N = \pm 1}
(i_{1} + \cdots + i_{N})^2  = N\ .
$$
This sum is equal to $N$ as can be easily proved e.g.\ by induction.
Our result in fact concerns not only case I but also case II because
their density operators coincide. Hence
$$\sigma^I_{N} = \sigma^{II}_{N} =\sqrt{N}$$.

All ensembles that can be obtained in the EPR experiment are
completely random, see, e.g., \cite{peres}, p.\ 148. Since
$\rho_N^{I} = \rho_N^{II}$, measurements of arbitrary $N$-particle
observables yield the same results for both preparations I and II.
We conclude that these preparations cannot be distinguished even as
ensembles of $N$-beams (for arbitrary $N$) and the superluminal
signals are not available.

\section{Correlations within a beam}
In the previous section, we have worked with purely random methods I
and II. However, if the two methods are modified so that
correlations are built in among particles of one beam, then
d'Espagnat's idea can be made use of. Let us formulate the idea so
that each appearance of paradox is removed. Indeed, d'Espagnat does
not claim that states described by the same statistical operator are
distinguishable (this would be the paradox). His idea is that one
and the same {\em ensemble} can be considered as representing a
one-particle state as well as an $N$-particle state, where $N$ is,
in principle, arbitrary. There can then be two different ensembles
(I and II) such that the corresponding one-particle states are
described by the same statistical operators while the corresponding
$N$-particle ones have different statistical operators. If the
measurements are restricted to one-particle quantities only, no
difference can of course ever be found between the two ensembles.
However, measurements of $N$-particle observables will find
differences. We are going to give some examples in this section.

In order to have definite correlations, we have to assume that the
states of individual particles within a beam are distinguishable and
can be labeled in a unique way. This can be achieved e.g.\ by
separating individual particle preparations by a suitable time
interval $\tau$, say, and utilizing the position degree of freedom
of the particle in addition to its spin. We assume that the
individual pure states are well-separated wave packets of the same
profile. This may still be considered as {\it one} definite
preparation method of a one-particle state, applied at different
times $t$, $t+\tau, \dots, t+N\tau, \dots, t+MN\tau, \dots$ and,
simultaneously, as a method of preparing an $N$-particle beam state
applied at times $t$, $t+N\tau, \dots, t+MN\tau, \dots$; $M$ and $N$
are positive integers. We have further to assume that the whole
arrangement leads to a dynamics that is time-translation invariant
so that it is sensible to speak of ``the same state'' at different
times.

As an example of a strict correlation, let us now additionally
specify method I (II) as follows: the $n$-th one-particle state is
prepared at the time $t+n\tau$ with spin component $\frac{1}{2}
(-1)^n$ into $Oz$ ($Ox$) axis.

If this is viewed as a preparation of a one-particle state,
 $N = 1$, then both resulting states are again described by
the same density matrices
$$
  \rho^I_1 = \rho^{II}_1 = \frac{1}{2}\left(\begin{array}{ll}1&0\\0&1
  \end{array} \right)\ .
$$
However, if considered as a preparation of a two-particle
state, $N = 2$, then the result is the pure state
 $e_{+1}(t)\otimes e_{-1}(t-\tau)$ for method I
and $f_{+1}(t)\otimes f_{-1}(t-\tau)$ for II. In this way, the two
preparation methods give two identical one-particle states (mixed)
but two {\em different} two-particle ones (pure). An easy
calculation gives
$$
\langle\Sigma_{z}\rangle_2^{I} =
\langle\Sigma_{z}\rangle_2^{II} = 0, \quad
\langle\Sigma_{z}^2\rangle_2^{I} = 0,
\langle\Sigma_{z}^2\rangle_2^{II} = 2,
$$
hence
$$
  \sigma^I_2 = 0\ ,\quad \sigma^{II}_2 = \sqrt{2}\ ,
$$
in agreement with \cite{Esp}.

Thus in this (rather special) case of preparation, the numbers that
agree with \cite{Esp} are obtained. Now consider the described
methods I and II as preparations of a three-particle state, $N = 3$.
Then, two different pure states are mixed with equal frequencies:
for example, in method I the states
$$
  e_{+1}(t)\otimes e_{-1}(t-\tau)\otimes e_{+1}(t-2\tau)
$$
and
$$
  e_{-1}(t)\otimes e_{+1}(t-\tau)\otimes e_{-1}(t-2\tau)
$$
are mixed with equal weights $\frac{1}{2}$. Clearly, the
three-particle beams can be described in $2^3 = 8$-dimensional
Hilbert space with basis $\{  e_{i_1}\otimes e_{i_2}\otimes e_{i_3}
\}_{i_k = \pm 1}$. The 8 basis vectors can be labeled, via the
correspondence $i_k = (-1)^{\alpha_k}$, by 3-digit  binary numbers
$(\alpha_1,\alpha_2,\alpha_3)$ in increasing order $0,1,\dots,7$.
Then the density matrix of the mixed 3-beam is diagonal,
$$
\rho_3^I = \rm{diag}(0,0,\frac{1}{2},0,0,\frac{1}{2},0,0),
$$
and
$$
\Sigma_z = \rm{diag}(3,1,1,-1,1,-1,-1,-3), \quad
\Sigma_z^2 = \rm{diag}(9,1,1,1,1,1,1,9).
$$
This gives
$$
\langle\Sigma_{z}\rangle_3^{I} =0, \qquad
\langle\Sigma_{z}^2\rangle_3^{I} = 1,
$$
hence
$$    \sigma_3^I = 1. $$

In method II we can work with the basis
$\{ f_{i_1}\otimes f_{i_2}\otimes
f_{i_3} \}_{i_k = \pm 1}$.
Then the density matrix has the same diagonal form
$$
\rho_3^I = \rm{diag}(0,0,\frac{1}{2},0,0,\frac{1}{2},0,0),
$$
but $\Sigma_z$ and $\Sigma_z^2$ are non-diagonal.
They can be explicitly found by using the action
$\sigma_z f_{\pm 1} = f_{\mp 1}$. In this way we obtain
$$
\langle\Sigma_{z}\rangle_3^{II} =0, \qquad
\langle\Sigma_{z}^2\rangle_3^{II} = 3,
$$
hence
$$    \sigma_3^{II} = \sqrt{3}. $$
We see that the 3-beams are also distinguishable.
However, the corresponding dispersions $\sigma^I_3$ and
 $\sigma^{II}_3$ do not agree
with those of \cite{Esp} (being both non-zero).

For general $N>3$ we have to distinguish the cases when
$N$ is even or odd. For even $N=2K$ we have a pure
$(2K)$-particle state with binary labels $0101\dots 01$
and obtain
$$
\langle\Sigma_{z}\rangle_{2K}^{I} =
\langle\Sigma_{z}\rangle_{2K}^{II} =0, \quad
\langle\Sigma_{z}^2\rangle_{2K}^{I} = 0,
 \quad \langle\Sigma_{z}^2\rangle_{2K}^{II} = 2K,
$$
hence
$$    \sigma_{2K}^{I} =0, \quad
\sigma_{2K}^{II} = \sqrt{2K}
 $$
in agreement with \cite{Esp}. For odd $N=2K+1$ we have a mixture of
two pure $(2K+1)$-particle states with $2K+1$ binary labels
$0101\dots 10$ and $1010\dots 01$, respectively, and equal weights
$\frac{1}{2}$. Straightforward computation now results in
$$
\langle\Sigma_{z}\rangle_{2K+1}^{I} =
\langle\Sigma_{z}\rangle_{2K+1}^{II} =0, \quad
\langle\Sigma_{z}^2\rangle_{2K+1}^{I} = 1, \quad
\langle\Sigma_{z}^2\rangle_{2K+1}^{II} = 2K+1,
$$
i.e.
$$    \sigma_{2K+1}^{I} =1, \qquad
 \sigma_{2K+1}^{II} = \sqrt{2K+1}.
 $$
We conclude that in the above examples the  $N$-beams contain more
information than the one-particle state prepared by the same
procedure.\footnote{Note that e.g. the beams with even $N=2K$ need
not consist of sequences of the special type $010101\ldots01$, but
any sequence will do the job provided that the numbers of 0's and
1's remain the same for all beams composing the ensemble. So the
interesting cases in which the physical ensembles prepared by
methods I and II are theoretically distinguishable are not as
special as the given examples would suggest \cite{Espc}.}

Our final remark concerns the {\it quantum statistics}. Since we are
working with particles of spin $1/2$, the states of an $N$-particle
beam must be antisymmetric in the {\em particle names}. In the
notation of tensor products (of states and operators), the particle
names are the order numbers of the factors in each such product
taken, say, from left to right. For fermions, these products must,
therefore, be antisymmetrized and provided by a normalization
factor. As an example, consider the two-particle state
$e_{+1}(t)\otimes e_{+1}(t-\tau)$. A fermion state obtained from it
is
$$
  \frac{1}{\sqrt{2}}\Big[e_{+1}(t_1)\otimes e_{+1}(t_2-\tau) -
  e_{+1}(t_2)\otimes e_{+1}(t_1-\tau)\Big]\ .
$$

In the previous section we could also work with one-particle states
distinguished by their times in the way described in this section.
For all our purposes, it has been satisfactory to consider only
tensor-product states containing one-particle ones in the order of
decreasing time, as in the above example. Then a bijective linear
mapping of our ordered states onto the antisymmetric fermion states
is well defined. This map preserves the number of states and their
inner products.  Since all our arguments are based on counting the
elements of state bases, they are preserved by this map.

It is also clear that everything could be done, in an
analogous manner, with photons, or with other discrete
degrees of freedom than spins.

\section{Concluding remarks}

Let us return to the quotations from \cite{Esp} in Sects. 1 and 2.
On pp. 122-123 of \cite{Esp} there follow the results on
$N=2K$-beams formed as random mixtures characterized by zero total
spin, or by fixed $m=K$ in our notation:

`Ensembles of these beams should therefore be considered. Such
ensembles may in principle be prepared and can therefore be
subjected to statistical measurements. For example, the fluctuations
of the quantity
$$ \Sigma_{z} = \sum_{n=1}^{N} \, \sigma_{z,n} $$
where  $\sigma_{z,n}$ is twice the $z$ component of the spin of the
$n$th particle in the beam and $N$ is the number of such particles,
can be experimentally measured on ensembles of beams prepared by any
of the two methods (I and II).' ... `We then observe that these
fluctuations are different in the two cases. In fact they are
characterized by the standard deviations
$$ \sigma^{I} = 0, \quad  \sigma^{II} = \sqrt{N}.$$
So that beams prepared by methods I and II can in principle be
distinguished from each other.'

In fact, such possibility was the reason for our query, whether
information could be transmitted in this way in the qualitatively
different context of the EPR experiment. Therefore in Sect. 3 we
considered an idea inspired from d'Espagnat's one \cite{Esp},
applied it to EPR situations and found that in these situations
information cannot be transmitted by using $N$-beams.

Finally we note that there is a general point of view implicit in
our paper. In Sect. 3 the ensemble of $N$-beams was defined by
giving very partial information about it and led to statistical
predictions to be verified by experiment. Then in Sect. 4 we defined
the ensemble of $N$-beams by giving more detailed information about
it, implying entanglement between the particles. The considered
cases can be compared with general situations in quantum mechanics
involving an ensemble of entangled particles in spatially separated
regions $A$ and $B$. Very partial information is obtained by
observing only the particles in region $A$, neglecting partners in
region $B$. It is well known that this situation is described by a
density matrix which is not a pure state. However, when the
information is obtained in both regions $A$ and $B$, then, because
of entanglement, the experimentalist finds results different from
the predictions in the former case.

\subsection*{Acknowledgements}
The authors would like to express their gratitude to Professor B.
d'Espagnat for careful reading of the manuscript and his valuable
comments which helped to improve the presentation. This research was
suppported by The Swiss National Foundation and by the Tomalla
Foundation, Z\"urich. J.T. acknowledges partial support by the
Ministry of Education of Czech Republic under the research project
MSM210000018.

\end{document}